\documentclass[lettersize,journal]{IEEEtran}
\usepackage{comment}
\usepackage{amsmath,amsfonts}
\usepackage{algorithmic}
\usepackage{algorithm}
\usepackage{array}
\usepackage[caption=false,font=normalsize,labelfont=sf,textfont=sf]{subfig}
\usepackage{textcomp}
\usepackage{stfloats}
\usepackage{url}
\usepackage{verbatim}
\usepackage{xcolor}
\usepackage{graphicx}
\usepackage{cite}
\usepackage[utf8]{inputenc}
\begin{document}

\title{Semi-Autonomous Prosthesis Control Empowered by 5G and Mobile Edge Computing}

\author{Ozan Karaali,~\IEEEmembership{Graduate Student Member,~IEEE,}
Hossam Farag,~\IEEEmembership{Member,~IEEE,}
Strahinja Do\v{s}en*,~\IEEEmembership{Member,~IEEE,}
and~\v{C}edomir Stefanovi\'{c}*,~\IEEEmembership{Senior Member,~IEEE}
\thanks{ This work was supported by the Independent Research Fund Denmark (DFF) through the CLIMB project no. 2035-00169B. (* \emph{equal contributions})}
\thanks{O. Karaali, H. Farag, and \v{C}. Stefanovi\'{c} are with the Department of Electronic Systems, Aalborg University, 9220 Aalborg, Denmark (e-mail: ozank@es.aau.dk; hmf@es.aau.dk; cs@es.aau.dk).}
\thanks{S. Do\v{s}en is with the Department of Health Science and Technology, Aalborg University, 9220 Aalborg, Denmark (e-mail: sdosen@hst.aau.dk).}}

\maketitle

\begin{abstract}
Prosthetic hands equipped with cameras can use computer vision to plan grasps automatically, reducing cognitive effort. However, running modern vision models on wearable devices is impractical due to power and processing constraints. We present the first prototype of a 5G-connected mobile edge computing (MEC)-enabled semi-autonomous prosthetic hand, which streams RGB-D images to an edge server for real-time grasp planning. Thirteen able-bodied participants performed pick-and-place tasks under six conditions: manual EMG control, on-device inference, wired Ethernet connectivity, and three 5G connectivity configurations (private 20~MHz network, private 100~MHz network, and a commercial 5G link) to the server. All network-based conditions performed similarly, achieving task times around 8.6~s (34\% faster than manual control), failure rates of 20--38\%, and 62\% lower overall workload. On-device processing performed worst with 10.3~s task time and 76\% failure rate due to slow embedded inference (3~fps vs.\ 6--20~fps over the network). Network latencies remained below 180~ms for private 5G and 270~ms for commercial 5G. All 5G configurations, including bandwidth-constrained and commercially variable networks, matched wired Ethernet performance while significantly outperforming both manual control and local processing, establishing 5G edge-offloading as a practical path to deploying compute-intensive prosthesis control.
\end{abstract}


\section{Introduction}

\IEEEPARstart{A}{pproximately} 2.3 million people in the United States live with limb loss, with upper extremity amputations accounting for 9.2\% of cases~\cite{rivera_estimating_2024}. Modern multi-articulated robotic prostheses can perform a variety of grasping patterns, but controlling them using surface electromyography (EMG) remains cognitively demanding~\cite{farina_extraction_2014}. In this approach, the electrical activity of the user's muscles is recorded using electrodes placed on the surface of the residual limb and translated into prosthesis commands. The users must generate distinct muscle activation patterns to produce explicit commands for each action: selecting grip type, adjusting wrist orientation, and modulating hand aperture, which can easily become exhausting when controlling devices with many degrees of freedom. Such cognitive burden contributes to inefficient control and device abandonment: Salminger et al.~\cite{salminger_current_2022a} reported rejection rates of 44\% for myoelectric prostheses, often citing difficulty of use.

Several approaches have been proposed to reduce this burden. The semi-autonomous control is a promising direction: in this scheme, the prosthesis is equipped with additional sensors that provide environmental information, allowing the prosthesis controller to perform some tasks automatically with minimal or no user intervention \cite{guo_humanintheloop_2023a}. For instance, cameras and/or depth sensors can be placed on the prosthesis and/or the user, and computer vision is used to perceive objects and adjust the hand configuration accordingly ~\cite{markovic_stereovision_2014, ghazaei_deep_2017, castro_continuous_2022}. The hand configuration parameters can be determined by estimating object properties (shape and size) using depth sensing and point cloud analysis~\cite{castro_continuous_2022}, or by directly classifying the acquired object images into associated grasp types using machine learning~\cite{ghazaei_deep_2017}. Therefore, the user's role is now shifted from low-level motor execution to high-level intent specification. The user decides \textit{what} to grasp and \textit{when}, while the system automatically handles \textit{how} by computing the appropriate grip type, wrist rotation, and aperture from visual input. Studies suggest that this shared-control approach can improve task performance while reducing mental and physical effort~\cite{mouchoux_artificial_2021}.


However, the computational requirements of real-time computer vision exceed the modest processing and data resources of compact wearable systems, such as prosthetic hands. Object detection and segmentation using deep neural networks demand GPU-class hardware that is not available in a prosthetic device. This creates a fundamental gap: implementing effective assistance to the user requires computational resources incompatible with wearable form factors. Most of the prototypes presented in the literature use dedicated lab computers tethered to the prosthesis. This shows the feasibility and benefits of semi-automatic control, but does not allow clinical translation outside the lab.

Importantly, the framework of Mobile Edge Computing (MEC) offers a resolution: the intensive perception tasks can be offloaded to nearby servers over wireless networks, thereby enabling prostheses to access powerful computing resources. Shatilov et al.~\cite{shatilov_using_2019} demonstrated the benefits of Cloud access: they offloaded EMG gesture classification from a 3D-printed prosthesis to a cloud server via a smartphone intermediary, achieving classification delays of 90--370~ms. Cloud offloading also reduced smartphone power consumption by 40\%, extending battery life to over 8 hours. However, their architecture introduced multiple network hops (Bluetooth to phone, WiFi to cloud), focused on gesture recognition rather than vision-based grasp planning, and did not rely on MEC.

MEC places computational resources closer to users than centralized cloud infrastructure~\cite{mach_mobile_2017, shi_edge_2016}, reducing round-trip latency. For prosthetic control, latency is critical: Farrell and Weir~\cite{farrell_optimal_2007} found that delays beyond 100--125~ms degrade perceived responsiveness in myoelectric systems. Fifth-generation wireless networks offer improved characteristics for latency-sensitive applications. The 5G New Radio specification targets user-plane latencies under 4~ms for ultra-reliable low-latency communication (URLLC), with peak data rates exceeding 10~Gbps~\cite{ahad_technologies_2020}. Practical deployments achieve 10--50~ms latency, which is still sufficient for many real-time applications. The use of 5G has been demonstrated in healthcare contexts, including remote surgery~\cite{lacy_5gassisted_2019}, but has not been exploited for closed-loop prosthesis control.

\begin{figure}[t]
\centering
\includegraphics[width=\linewidth]{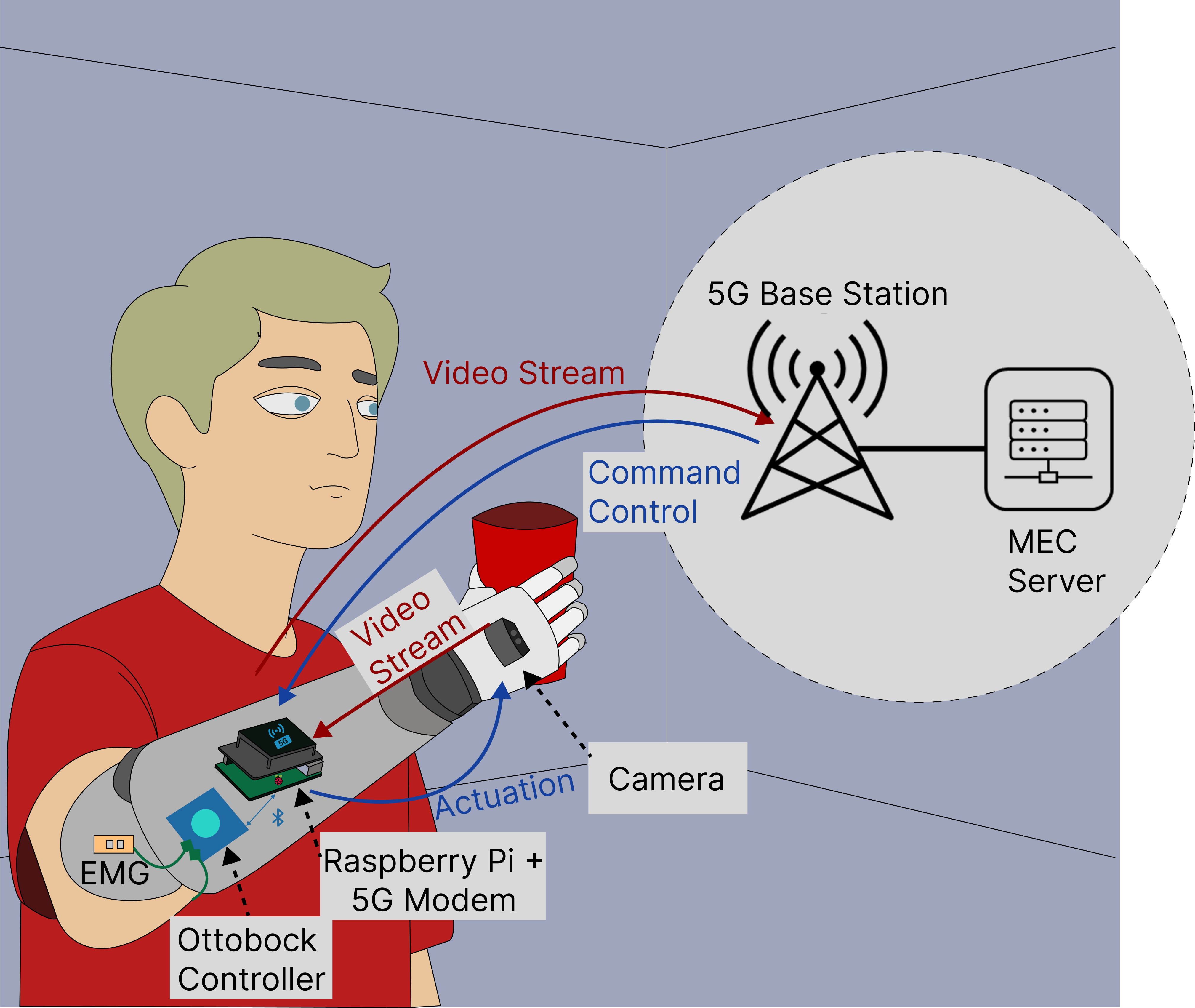}
\caption{The developed 5G and MEC-enabled prosthesis prototype. The camera mounted on the prosthetic hand records an object that the user wants to grasp. The recorded RGB-D frames are streamed to an edge server over 5G, which uses advanced computer vision to estimate hand configuration appropriate for grasping the object. The edge server returns grasp commands to the hand for automatic hand preshaping.}
\label{fig:concept}
\end{figure}

In~\cite{chiariotti_future_2024}, the concept of \textit{connected bionic limbs} was proposed, in which advanced perception and real-time decision-making are enabled by offloading computationally intensive tasks from wearable prosthetic devices to nearby edge computing resources via next-generation wireless connectivity. By providing bionic limbs with access to Edge and Cloud, this approach can turn robotic prostheses into next-generation, highly flexible and powerful systems able to leverage advanced control strategies that are normally too computationally demanding for onboard execution. As explained above, 5G networks are particularly well-suited for implementing this paradigm~\cite{ahad_technologies_2020}. Building on this vision, preliminary studies~\cite{karaali_5genabled_2025, karaali_enabling_2025} systematically characterized the communication performance of private and commercial 5G networks for prosthetic applications, quantifying achievable latency, throughput, and variability under realistic deployment conditions. While these earlier works established the technical feasibility of 5G-based connectivity for prosthetic systems, they did not address the central question: Can 5G-based edge offloading support real-time, semi-autonomous prosthetic control with human users, achieving task performance and usability comparable to wired connectivity and better than the conventional, purely manual approach?

To answer this question, in this work, we implement and experimentally evaluate the first prototype of a semi-autonomous prosthetic hand that uses a 5G modem to connect directly to the control unit (mobile base edge server), without intermediate network hops. We conducted a comprehensive assessment of the novel system by conducting experiments involving thirteen participants performing functional grasping tasks using a prosthesis under six conditions: manual EMG control, on-device inference, wired Ethernet connection to the edge server, and three 5G configurations, namely, private 20~MHz, private 100~MHz, and commercial 5G connection, where the former two connect to the edge server and the third effectively creates a connection to the cloud-server. For each experiment, the performance is assessed by measuring task completion time, failure rate, and subjective workload.

Our results show that responsive and efficient 5G-connected prosthesis control is not only viable but advantageous. All wireless network-based conditions significantly outperformed manual control. Critically, all 5G configurations, including bandwidth-limited deployments (e.g., the 20 MHz link) and the commercial network connection that provides no bandwidth guarantees, achieved equivalent performance to that of the wired Ethernet. On-device processing, despite avoiding network latency entirely, performed worst due to insufficient inference speed. These findings demonstrate that 5G-connected prostheses can work in real-world conditions and outperform current clinical approaches.

\section{5G-connected MEC-enabled Prosthesis Prototype}

\subsection{System components}

\begin{figure*}[t]
\centering
\centerline{\includegraphics[width=0.85\linewidth]{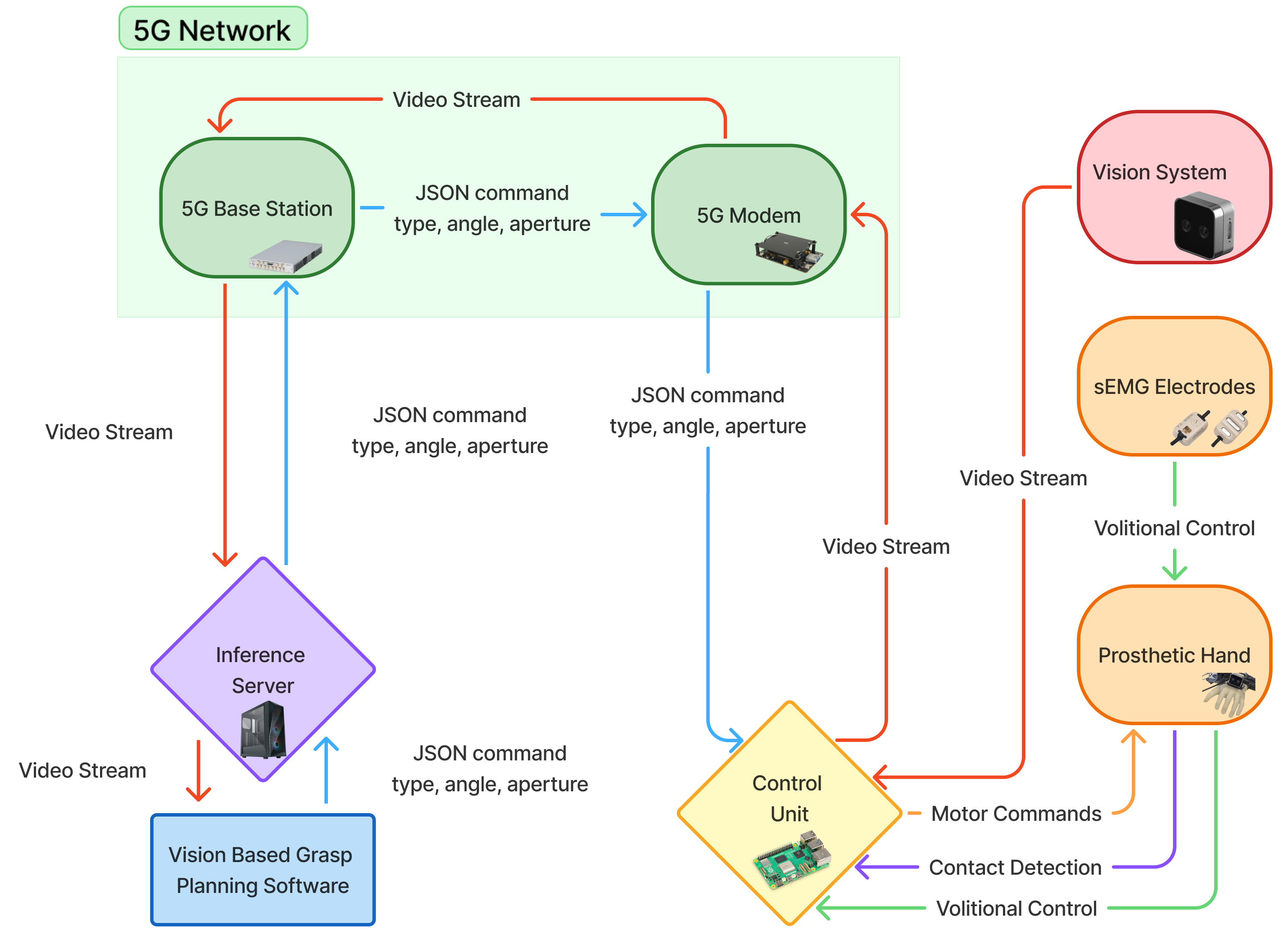}}
\caption{Detailed system architecture and data flow in the developed prototype. The camera placed on the prosthesis streams the images to the control unit. The control unit sends the frames via a 5G modem to the 5G base station setup in the lab, where an inference server performs computer vision processing for grasp planning. The commands are sent back via 5G to the control unit, which transmits them to the prosthesis. EMG electrodes are placed on the forearm of the user to enable volitional control (semi-automatic scheme). The user employs EMG (muscle contractions) to trigger automatic control and to orient the prosthesis manually. The details are explained in the text.}
\label{fig:sys_arch}
\end{figure*}

The developed prototype is depicted in Fig.~\ref{fig:concept}, and the data flow between the different components is provided in detail in Fig.~\ref{fig:sys_arch}. The prototype includes the following components.
\subsubsection{Prosthetic Hand}
The Ottobock Michelangelo hand is a multi-articulated myoelectric prosthesis with powered fingers and a repositionable thumb, capable of palmar (power) and lateral (key) grasp patterns~\cite{ottobock_michelangelo}. Two Ottobock 13E200 surface EMG electrodes~\cite{ottobock_electrode_13e200}, positioned over the forearm flexor and extensor muscles, connect to the built-in prosthesis proprietary controller, which samples the EMG signals at 100~Hz. The hand exposes a Bluetooth interface through which the control unit (Raspberry Pi~5) reads EMG data and sends motor commands, effectively overriding the default firmware control. Built-in force sensor detects object contact and measures grasping force. A wrist rotation motor provides powered pronation/supination, with software-limited range of $-59^\circ$ to $+120^\circ$ to prevent unnatural positions. The prosthesis uses its own built-in battery.

\subsubsection{Vision System}
An Intel RealSense D405 RGB-D camera~\cite{intel_realsense_d405}
was mounted on the dorsal surface of the prosthetic hand and oriented to capture the workspace in front of the prosthesis. The D405 provides synchronized color and depth imagery at the resolution of 640$\times$480 pixels and frame rate of 30~fps, with an optimal operating range of 7--50~cm and 87$^\circ$$\times$58$^\circ$ field of view. This near-field capability is particularly suitable for our application, namely, the detection and classification of objects placed close to the hand.

\subsubsection{Control Unit}
A Raspberry Pi 5 with 8~GB RAM~\cite{raspberrypi_5} serves as the local control unit, handling EMG acquisition, camera streaming, motor commands, and network communication. The local control unit is connected to a Quectel RM502Q-AE modem~\cite{quectel_rm502q}
via USB, providing direct 5G cellular connectivity without intermediate WiFi or router hops. A 20,000~mAh USB battery powers the control unit and modem, providing approximately 4--5 hours of continuous operation. 

\subsubsection{5G Private Network}
The network uses an Ettus USRP N310~\cite{ettus_usrp_n310}
software-defined radio as a base station, with 5G core functions virtualized via srsRAN~\cite{srsran_project} and Open5GS~\cite{open5gs} on the same edge server running computer vision and grasp inference pipeline for prosthesis control.

\subsubsection{Inference Server}
The edge server uses an AMD Ryzen 9 7950X processor~\cite{amd_ryzen_7950x} with 64~GB RAM and an NVIDIA RTX 5080 GPU\cite{nvidia_rtx_5080}. The server also hosts the 5G core network functions, minimizing routing latency.

\subsection{System operation}
\label{sec:sysop}

The system functions as a state machine with two states that correspond to automatic and manual control modes. Initially, the hand is in the \emph{Automatic control state}. The frames acquired by the camera are continuously streamed to the edge server over a 5G network (see Fig.~\ref{fig:sys_arch}) via WebSocket~\cite{fette_websocket_2011} using pipelined transmission with multiple concurrent in-flight requests. The effective frame rate is not explicitly capped but adapts to the achievable round-trip time and server processing speed, yielding approximately 20~fps over Ethernet and 6--16~fps over 5G, depending on the configuration (see Section~\ref{sec:results_network}). When the network cannot keep up, the most recent frame replaces older queued frames, ensuring the server always processes the latest camera view. Color and depth images are compressed as JPEG and 16-bit PNG, respectively. An adaptive compression mechanism monitors the round-trip time and adjusts the JPEG quality (range 30--90) to help sustain the frame rate under variable network conditions.

At the edge server, the frames are analyzed by the vision pipeline, explained in detail in Section~\ref{sec:vp}, to estimate object size and orientation. Based on this, the server computes the desired grasp parameters (grasp type, hand aperture, and wrist rotation angle) and sends them back over the network to the hand in the form of a JSON command (see Fig.~\ref{fig:sys_arch}) when it detects valid objects in the workspace. To grasp an object, the participant brings the hand in front of the object to place it in the camera's field of view. They then briefly activate their wrist flexor muscles and, in response, the hand automatically preshapes according to the latest grasp parameters received from the edge server. Specifically, the wrist rotates to align with the object orientation, the thumb repositions for the selected grasp type (palmar or lateral), and the hand opens to the computed aperture to match the object size.



After preshaping, the hand transitions into the \emph{Manual state}, in which the participant has full volitional control. The EMG signals captured by the electrodes are read by the control unit via Bluetooth (see Fig.~\ref{fig:sys_arch}) and translated into motor commands, as explained below. If the participant is satisfied with the automatically selected hand configuration, they simply activate the wrist flexor to close the hand around the object. If the automatic preshape is incorrect, however, the participant can reconfigure the hand using direct proportional EMG control. The wrist flexor and extensor muscles drive the prosthesis proportionally, and muscle co-contraction cycles the active function through wrist rotation, palmar grasping, and lateral grasping. The stronger the contraction, the faster the prosthesis moves. Such a 2-channel proportional control with switching is the standard clinical approach for prosthesis control. Once the hand opens and the object is released, in the next co-contraction, the system switches back to the Automatic state.

\begin{figure}[t]
\centering
\includegraphics[width=0.8\linewidth]{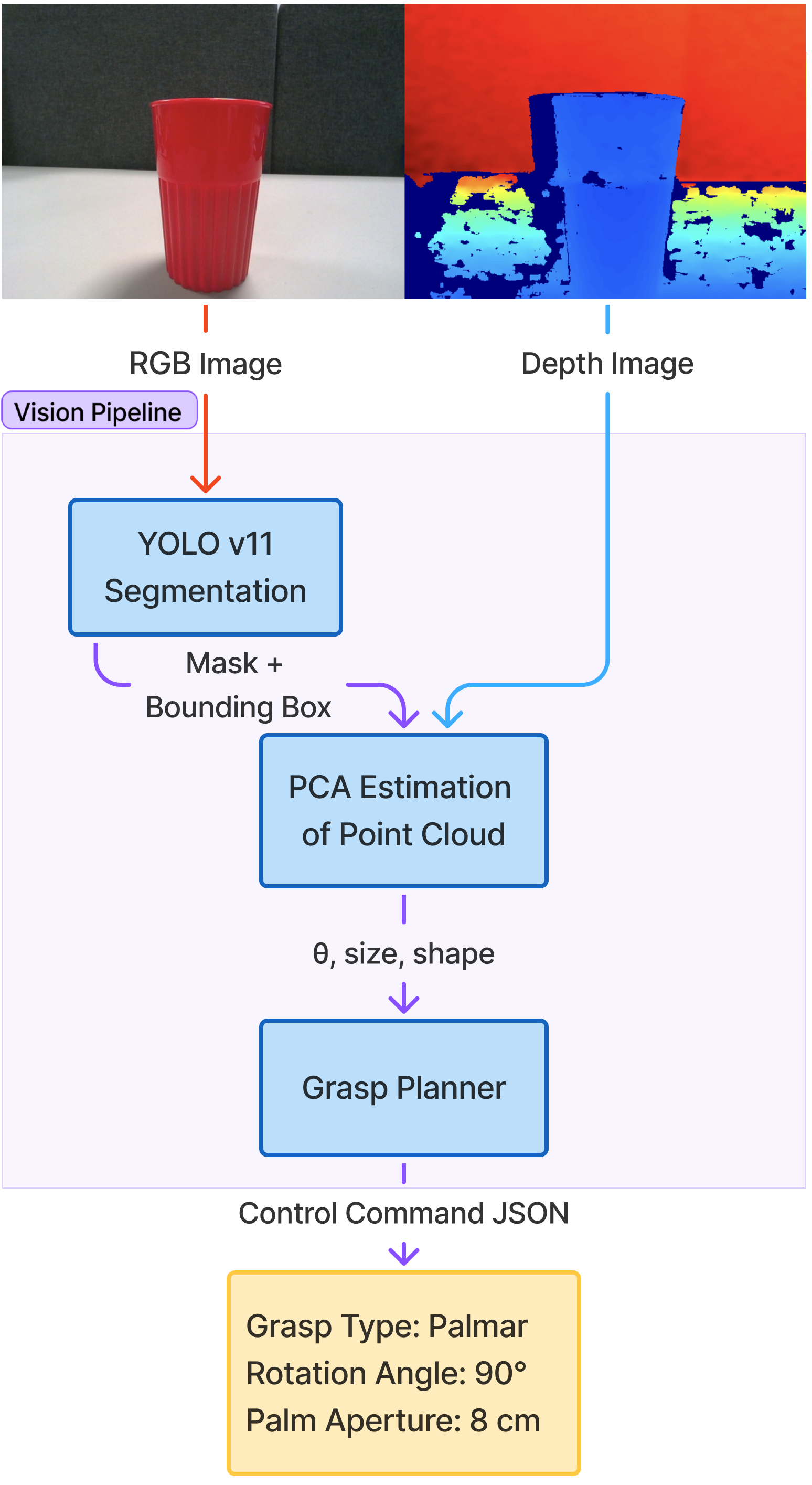}
\caption{Vision pipeline. Each RGB-D frame is processed by YOLOv11 for segmentation, followed by PCA-based estimation of object dimensions and orientation, and rule-based grasp planning that outputs grasp type, wrist rotation angle, and palm aperture appropriate for grasping the object recorded by the camera. The details are explained in the text.}
\label{fig:pipeline}
\end{figure}

\subsection{Computer Vision and Grasp Selection Pipeline}
\label{sec:vp}

The inference server processes each received camera frame through the pipeline shown in Fig.~\ref{fig:pipeline}. This pipeline is selected as a representative example but can be replaced by any other machine-learning or point-cloud processing approach presented in the literature~\cite{cirelli_computer_2026}.
Object detection and segmentation are performed using the YOLOv11-large model variant with instance segmentation (yolo11l-seg)~\cite{khanam_yolov11_2024}, optimized via TensorRT for GPU inference at the input resolution of 896$\times$896 pixels. The model uses weights pretrained on the COCO dataset~\cite{lin_microsoft_2014}, supporting 80 object classes. 

Detected objects are filtered by depth validity. Objects outside the camera's reliable depth range (7--50~cm) or with invalid depth readings are excluded. Principal component analysis (PCA) is performed on each segmentation mask to yield the object orientation, dimensions (length, width in cm), and centroid. The variance ratio is computed to classify the shape as round ($<$1.8) or elongated ($>$3.0).

Grasp type selection is implemented using a rule-based decision tree (similar to~\cite{castro_continuous_2022}). The geometry-based rules use the PCA-derived object dimensions as follows: if the object height exceeds 12~cm, a palmar grasp is selected; if the object width is less than 5~cm, a lateral grasp is chosen; for remaining objects, orientation angle determines the selection (lateral below 45$^\circ$, palmar otherwise). These geometric rules are augmented by class-specific overrides based on the YOLO-recognized object category (e.g., lateral for utensils and writing instruments, palmar for round fruits), and aspect-ratio heuristics to detect thin, elongated shapes regardless of classification, and a minimum thickness check to avoid lateral grasps on flat-lying objects. The wrist rotation angle is calculated from the PCA main axis with table-based orientation correction to avoid palm-up positions, and the palm aperture is set to the measured object width plus 1~cm of extra clearance, bounded by the limits of prosthesis range of motion (max aperture).

The estimated prosthesis configuration parameters are considered valid when: (1) YOLO successfully segments an object, (2) the object lies within the depth operating range, and (3) the required aperture falls within the prosthesis limits (2--12~cm for the palmar and 2--7.5~cm for the lateral grasp). If the current frame does not yield a valid detection, the system briefly reuses the most recent valid result as a short-term cache; if no cached result exists, the system does not invoke grasping and continues streaming until a valid detection is obtained.

\section{Experimental assessment}

\subsection{Tested Network Configurations}

We tested the performance of prosthesis use in six conditions, listed in Table~\ref{tab:conditions}, to perform a systematic analysis of how network characteristics affect prosthesis control performance.

\begin{table}[htbp]
\caption{Experimental Network Conditions}
\begin{center}
\begin{tabular}{|l|l|p{4cm}|}
\hline
\textbf{Code} & \textbf{Condition} & \textbf{Rationale} \\
\hline
MC & Manual Control & Clinical baseline: standard myoelectric control without AI \\
\hline
ODML & On-Device Inference & Fully wearable: no network dependency \\
\hline
EC & Ethernet & Upper bound: minimal network latency \\
\hline
5G-20 & Private 5G 20~MHz & Bandwidth-constrained: shared spectrum scenario \\
\hline
5G-100 & Private 5G 100~MHz & High-capacity: dedicated spectrum \\
\hline
5G-PUB & Commercial 5G & Public commercial network 5G link with variable (uncontrolled) bandwidth\\
\hline
\end{tabular}
\label{tab:conditions}
\end{center}
\end{table}

\emph{Manual Control (MC)} serves as the baseline, representing current state-of-the-art in clinical myoelectric control. In this condition, the automatic state is deactivated and, therefore, the participant controls the prosthesis manually, using a 2-channel direct and proportional scheme with switching, as explained in Section \ref{sec:sysop}. Comparing semi-autonomous conditions against this baseline quantifies the benefits of a smart connected prosthesis.

\emph{On-Device Inference (ODML)} represents fully wearable processing with no network dependency. In this condition, the entire vision pipeline, described in Section \ref{sec:vp}, runs on the local controller (Raspberry Pi 5) ~
using YOLOv11-nano~\cite{khanam_yolov11_2024}. This condition tests whether embedded processing with a simpler object detection and segmentation model, suitable for deployment on a wearable device, can match network-offloaded performance, using the full model.

\emph{Ethernet (EC)} provides a baseline for network-based performance, where a Gigabit wired direct connection to the server establishes a lower bound on the latency of wireless connectivity conditions.

\emph{Private 5G} conditions use our 5G-connected prosthesis prototype (standalone mode, SA) operating in Band n77, specifically, in the frequency range 4.1--4.2~GHz. We tested two bandwidth allocations:
\begin{itemize}
    \item \emph{20~MHz (5G-20):} Corresponds to a default configuration option, where the available bandwidth is shared among users.
    \item \emph{100~MHz (5G-100):} Represents a dedicated high-capacity deployment with exclusive spectrum access.
\end{itemize}

\emph{Commercial 5G Network (5G-PUB)} uses our 5G-connected prosthesis prototype operating over a public commercial 5G network (non-standalone, NSA). Traffic traverses the public internet between the cellular network and our server, thus effectively positioning the server in the cloud with respect to the prosthetic device. This condition reflects a deployment with shared spectrum, variable load in terms of the number of users being served by the base station and the network, and multi-hop routing through the network core.

\subsection{Participants}
We recruited 13 able-bodied participants ($N=13$; 10 male, 3 female; age range 24--63 years, $M=33.7$, $SD=11.4$). 
The inclusion criteria were normal or corrected-to-normal vision and no prior experience with myoelectric prosthesis. The experiments were approved by the \textit{[redacted for anonymity]} (protocol number \textit{[redacted]}). All subjects signed an informed consent form before commencing with the test.

\subsection{Task and Procedure}

\begin{figure*}[t]
\centering
\includegraphics[width=\linewidth]{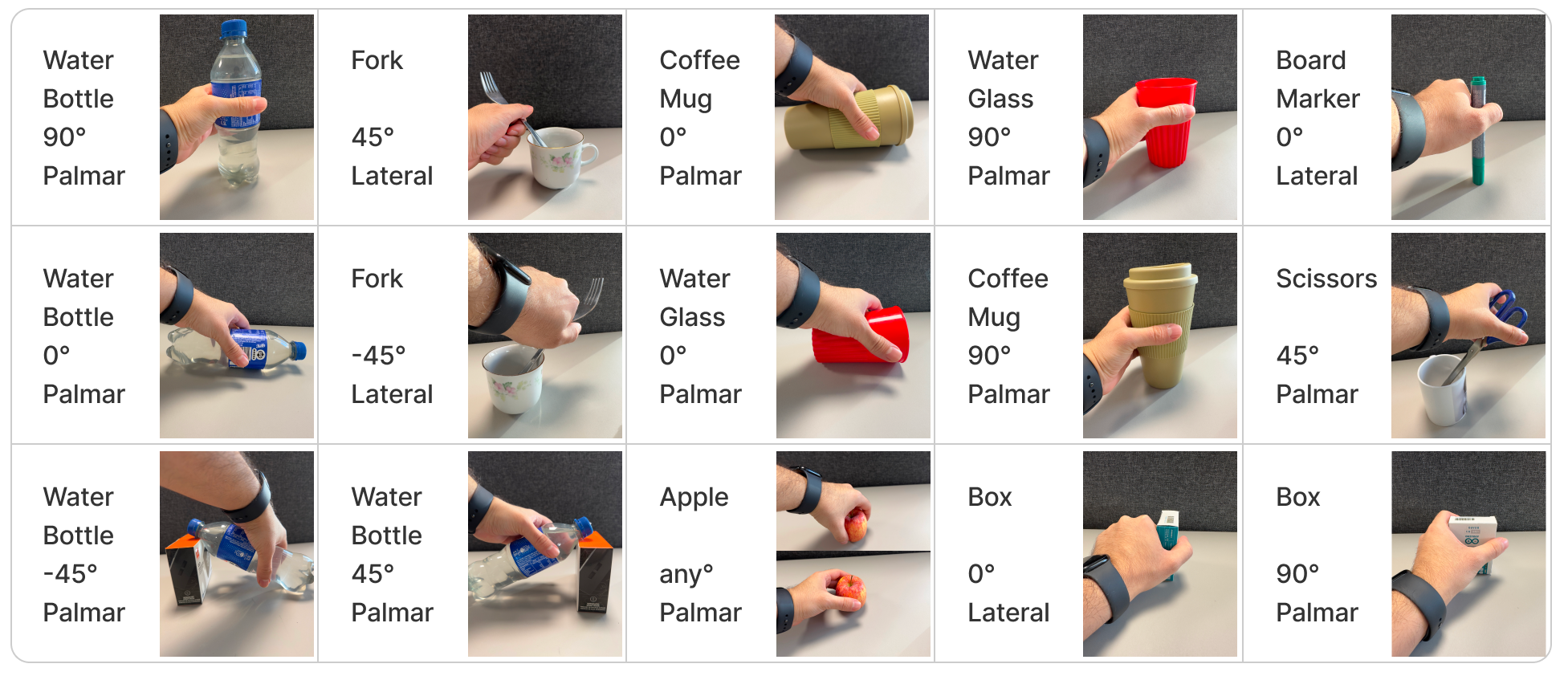}
\caption{Visual cues presented to participants. Each row shows an object at the tested orientations, with the required grasp type (palmar or lateral) indicated. In each condition, the participant performed 15 trials by following the object and orientation combinations shown above, from the top left to the bottom right panel.}
\label{fig:cues}
\end{figure*}

The participants used a prosthesis to perform a pick-and-place task. At the beginning of each trial, a visual cue displayed on a screen (Fig. \ref{fig:cues}) indicated to the participant the target object, the required grasp type (palmar or lateral), and the desired wrist rotation.

\begin{figure}[t]
\centering
\includegraphics[width=\linewidth]{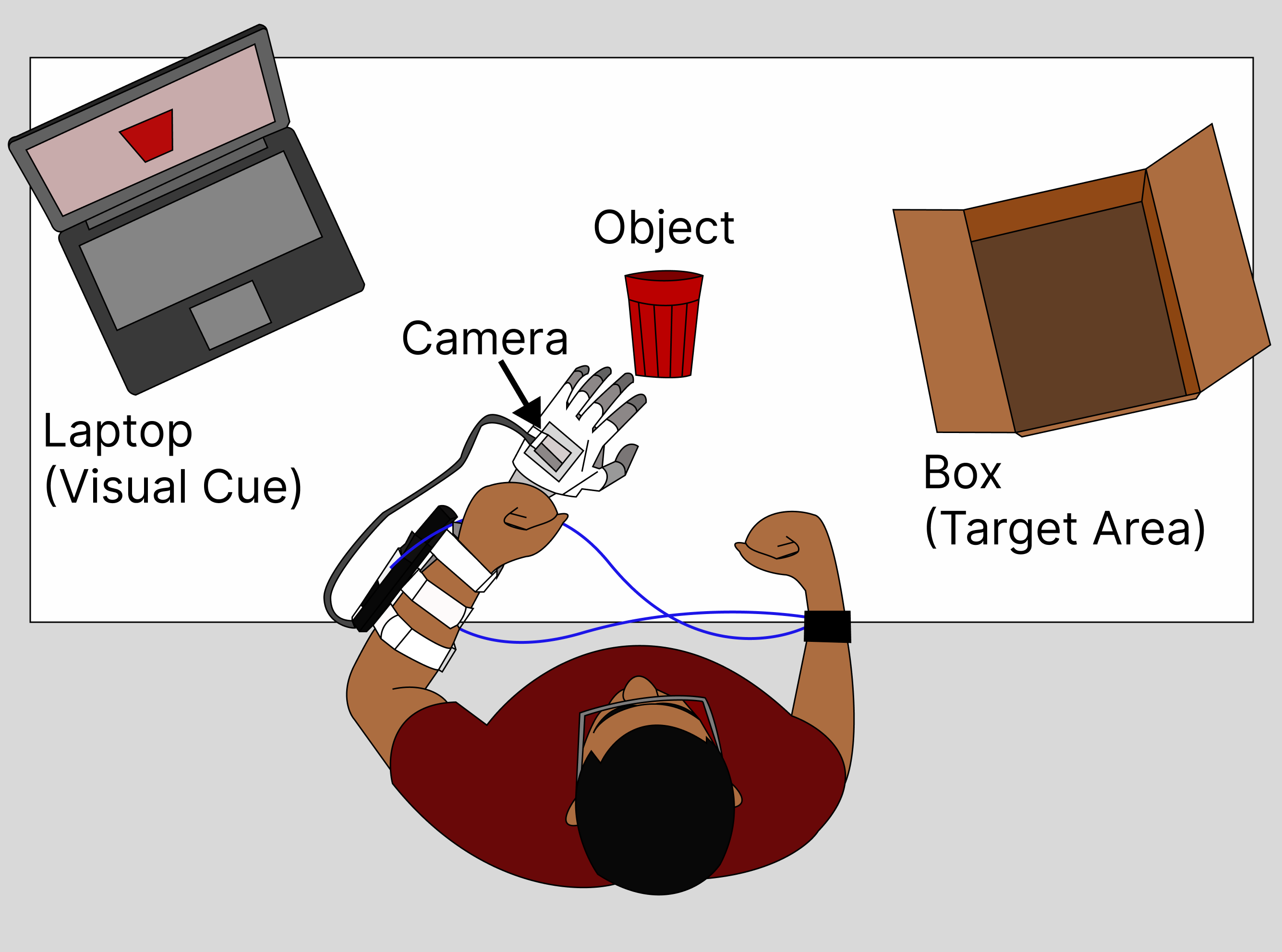}
\caption{Experimental setup (top view). The participant sits facing the table with a laptop displaying visual cues on the left, the target object at a marked position in the center, and a collection box on the right. The task for the participant was to grasp the object and replace it into the collection box.}
\label{fig:setup}
\end{figure}

The experimental setup is shown in Fig.~\ref{fig:setup}, and the task sequence was the following:
\begin{enumerate}
    \item The experimenter positioned an object at a marked location on the table placed in front of the participant.
    \item The participant approached the object, triggered the automatic hand preshaping, and if happy with the selected grasp, they closed the hand around the object. Otherwise, they manually reconfigured the prosthesis and then grasped the object. 
    \item They transported the object to the collection box and released it inside the box.
\end{enumerate}

A trial was considered successful if the participant grasped the object using the correct grasp type and wrist orientation and placed it inside the collection box. If the automatic preshape was incorrect, the participant could either manually correct the hand configuration using EMG control (as described in Section~\ref{sec:sysop}) or restart the trial. In case of failure, the participant could retry up to three times; if all attempts failed, the trial was skipped. A trial in which manual correction was needed was still counted as a failure of the automatic system.

The object set shown in Fig.~\ref{fig:cues}, included 8 common daily life items: a water bottle, a water glass, a coffee cup, a fork, a board marker, scissors, an apple, and a small box. Each object was presented in different orientations, namely, $0^\circ$, $\pm45^\circ$, and $90^\circ$ relative to the participant midline, challenging the system with a range of wrist rotation demands that can be encountered during everyday reaching and grasping.

The participants performed 15 trials (one per object orientation combination) in a fixed sequence, as shown in (Fig.~\ref{fig:cues}), in each of the six conditions listed in the previous section. For each trial, the participant had up to three attempts to complete the grasp; a trial was marked as failed if all three attempts were unsuccessful or if the trial timed out. The order of experimental conditions (see Table~\ref{tab:conditions}) was counterbalanced across participants using a Latin Square design~\cite{grant_latin_1948}.

Before the data collection, each participant received a 5-minute verbal introduction to myoelectric control and prosthesis operation. Then, the EMG electrodes were placed on their forearm and calibrated by recording maximum voluntary flexion and extension to set the proportional control range. This was followed by a single 10-minute familiarization session at the beginning of the experiment, during which participants were introduced to both control modes: manual proportional EMG control with co-contraction mode switching, and semi-autonomous control with the triggering of automatic preshaping. The participants were instructed to position the hand so that the camera could see the target object before triggering the automatic control. The sessions lasted approximately 2 hours per participant, with 15--20 minutes per condition, including rest breaks between the conditions and trials as requested by the participant.

\subsection{Outcome Measures}
The primary metrics to assess the performance across conditions were:
\begin{itemize}
    \item \emph{Task completion time} $T_c$, measured as the time elapsed from cue presentation (trial start) to object release in the collection box.
    \item \emph{Failure rate $F_r$} was defined as the percentage of trials with rotation errors ($>$30$^\circ$ deviation from the target angle), incorrect grasp type selection, grasp failures (when the hand could not acquire the object), or drops (the object falls out from the hand during transport).
    \item \emph{Subjective workload}, measured using the NASA Task Load Index (NASA-TLX)~\cite{HART1988139}, which is a standard measure to assess perceived workload across six dimensions: mental demand, physical demand, temporal demand, performance, effort, and frustration. Each dimension is rated on a 0--100 scale in increments of 10. We used raw (unweighted) scoring, averaging the six subscales without participant-derived weights, as raw TLX has been shown to maintain sensitivity comparable to the weighted version~\cite{HFES_2006_Paper}. After each condition, the participants rated the six subscales directly, and their scores were reviewed together with the experimenter to ensure the ratings reflected their intended assessment.
\end{itemize}

We also continuously logged the following technical parameters to characterize the connectivity performance (except for the ODML condition):
\begin{itemize}
    \item Round-trip latency via WebSocket ping timestamps.
    \item Effective bandwidth estimated from bytes transferred per second.
    \item Inference frame rate computed as the number of detections received per second.
    \item GPU processing time logged on the server-side.
\end{itemize}

\subsection{Statistical Analysis}

We assessed the normality of each outcome measure per condition using the Shapiro-Wilk test. Task completion times violated the normality assumption in three of six conditions ($p < 0.05$). Although failure rates and NASA-TLX scores did not violate normality per condition, we adopted non-parametric tests uniformly across all measures for consistency and because the small sample size ($N=13$) limits the power of normality tests. Specifically, we used: (i) the Friedman test for omnibus comparisons across conditions, and (ii) Wilcoxon signed-rank tests for post-hoc pairwise comparisons, with Bonferroni correction.

\section{Results}

\subsection{Network Performance}
\label{sec:results_network}
Table~\ref{tab:network} summarizes network performance metrics across all conditions with semi-autonomous control. The mean GPU processing time on the edge server was approximately 22~ms, while on-device inference on a Raspberry Pi took on average 429~ms, i.e., roughly 20 times longer.

This processing advantage compensated for network transmission delays: despite ODML entirely avoiding network latency, its slow inference resulted in overall worse responsiveness than all network-connected conditions. Commercial 5G (median RTT 234~ms, total latency 270~ms) achieved an inference frame rate of 6.1~fps. Private 5G performed better: 5G-100 reached 16.4~fps with 94~ms total latency, approaching wired Ethernet (19.9~fps, 35~ms). 5G-20 achieved 11.0~fps at the latency of 177~ms.

Tail latencies (P95) were 287--472~ms for private networks and 890~ms for commercial 5G. ODML exhibited severe tail latency (P95: 2249~ms), despite a median inference time of 429~ms. This variability is attributed to CPU contention among concurrent tasks (inference, camera capture, EMG processing) and the resulting frame dropping, as the pipeline could sustain only approximately 3~fps.
%

\begin{table}[htbp]
\caption{Network Performance Metrics}
\begin{center}
\footnotesize
\begin{tabular}{|l|c|c|c|c|}
\hline
\textbf{Condition} & \textbf{RTT (ms)} & \textbf{Total (ms)} & \textbf{P95 (ms)} & \textbf{FPS} \\
\hline
ODML & 104* & 596 & 2249 & 3.0 \\
EC & 10 & 35 & 289 & 19.9 \\
5G-20 & 144 & 177 & 472 & 11.0 \\
5G-100 & 65 & 94 & 287 & 16.4 \\
5G-PUB & 234 & 270 & 890 & 6.1 \\
\hline
\end{tabular}
\label{tab:network}
\end{center}
\footnotesize{RTT = network round-trip time (median). Total = RTT + processing. GPU processing $\approx$ 22~ms; ODML processing $\approx$ 429~ms. *For ODML, RTT represents local inter-process communication delay.}
\end{table}

\subsection{Task Performance}

Friedman tests revealed significant effects of condition on all primary metrics: task completion time ($p < 0.0001$), failure rate ($p < 0.0001$), and NASA-TLX workload ($p < 0.0001$).

\begin{figure*}[t]
\centering
\subfloat[Task Completion Time\label{fig:results_time}]{\includegraphics[width=0.48\linewidth]{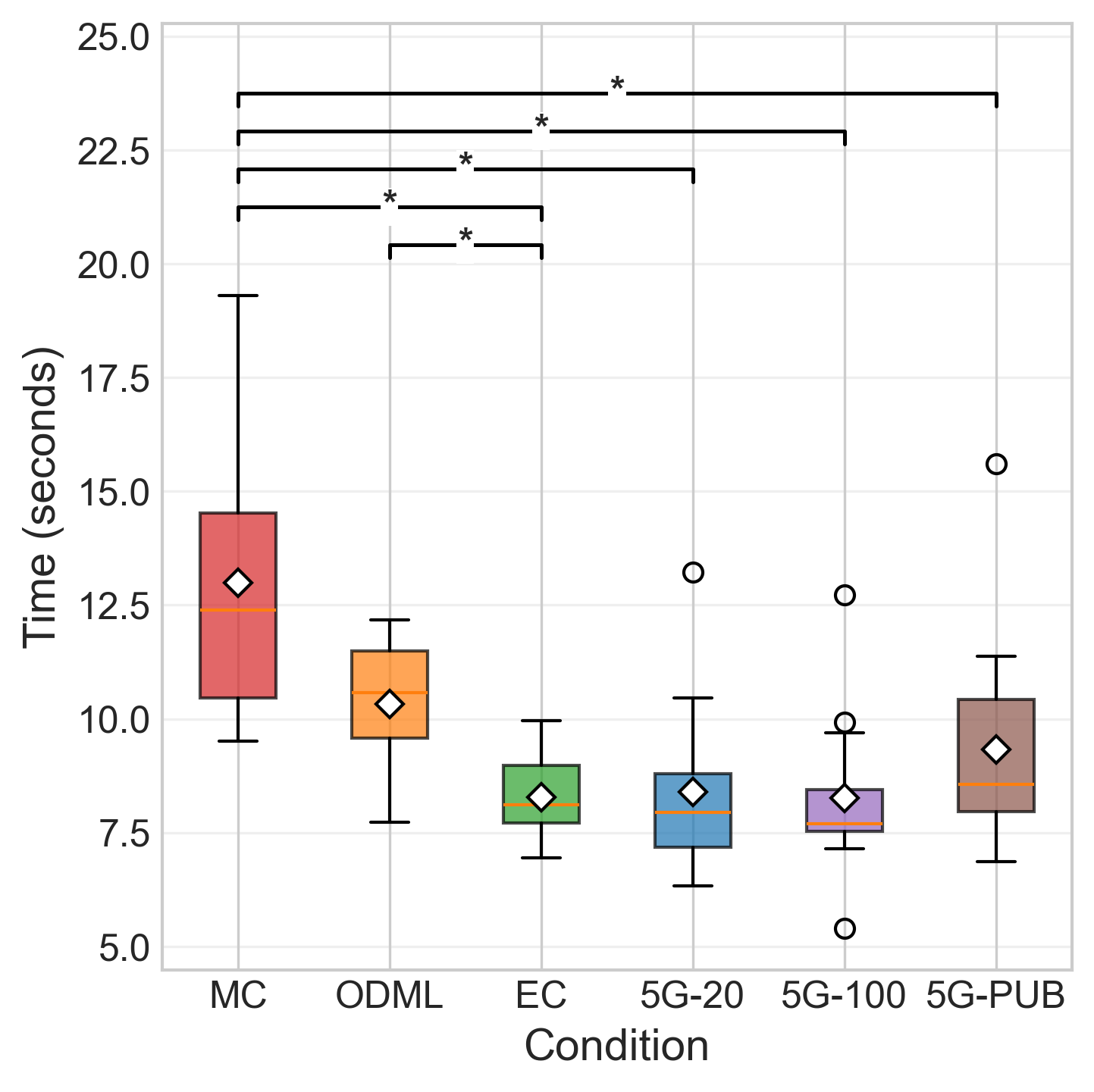}}
\hfill
\subfloat[Failure Rate\label{fig:results_failure}]{\includegraphics[width=0.48\linewidth]{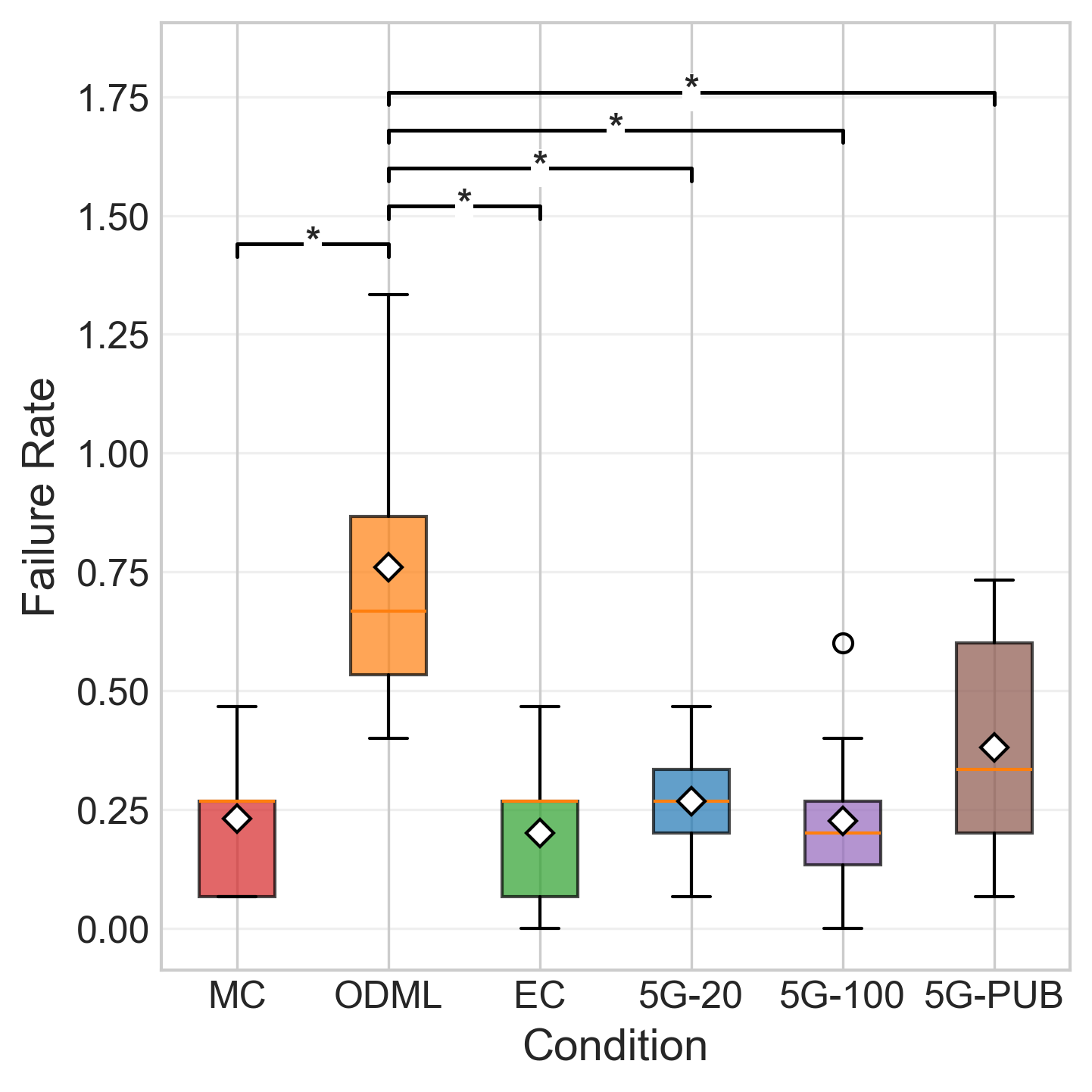}}
\caption{Objective task performance for different control conditions: (a) completion time and (b) failure rate. Box: interquartile range (IQR, 25th--75th percentile); horizontal line shows the median value; $\diamond$ shows the mean value; whiskers show 1.5$\times$IQR; dots represent the outliers. Brackets with $*$ denote statistically significant pairwise differences ($p<0.05$, Wilcoxon signed-rank test with Bonferroni correction).}
\label{fig:results_objective}
\end{figure*}

The post-hoc Wilcoxon signed-rank tests revealed that all connected prosthesis conditions yielded significantly shorter task completion times compared to manual control (all $p_{adj} < 0.03$; Fig.~\ref{fig:results_time}). The mean completion times decreased from 12.99~s under manual control to approximately 8.27~s and 8.26~s under Ethernet and 5G-100, respectively, corresponding to a relative reduction of about 36\%. Commercial 5G also significantly outperformed manual control ($p_{adj} = 0.028$), confirming that the proposed 5G-connected prosthesis framework retains performance benefits under real-world network conditions.

Failure rates under connected prosthesis conditions ranged between 20--38\% (Fig.~\ref{fig:results_failure}) and were not significantly different from those achieved with manual control ($p_{adj} > 0.05$), confirming that the observed reductions in task completion time were not achieved at the expense of task success. On the other hand, on-device inference (ODML) achieved the highest failure rate (75.9\%), significantly worse than all connected prosthesis conditions (all $p_{adj} < 0.035$), and even manual control (23.1\% failure rate, $p_{adj} = 0.033$). These results confirm that computation offloading is necessary for robust task-level performance: the computational advantage of GPU servers outweighs the latency cost of 5G-network transmission. 

Pairwise comparisons between Ethernet, private 5G, and commercial 5G conditions revealed no statistically significant differences across task time, failure rate, and workload ($p_{adj} > 0.05$).

\subsection{Subjective Workload}

\begin{figure*}[t]
\centering
\subfloat[NASA-TLX Overall Score\label{fig:results_tlx}]{\raisebox{0.02\linewidth}{\includegraphics[width=0.48\linewidth]{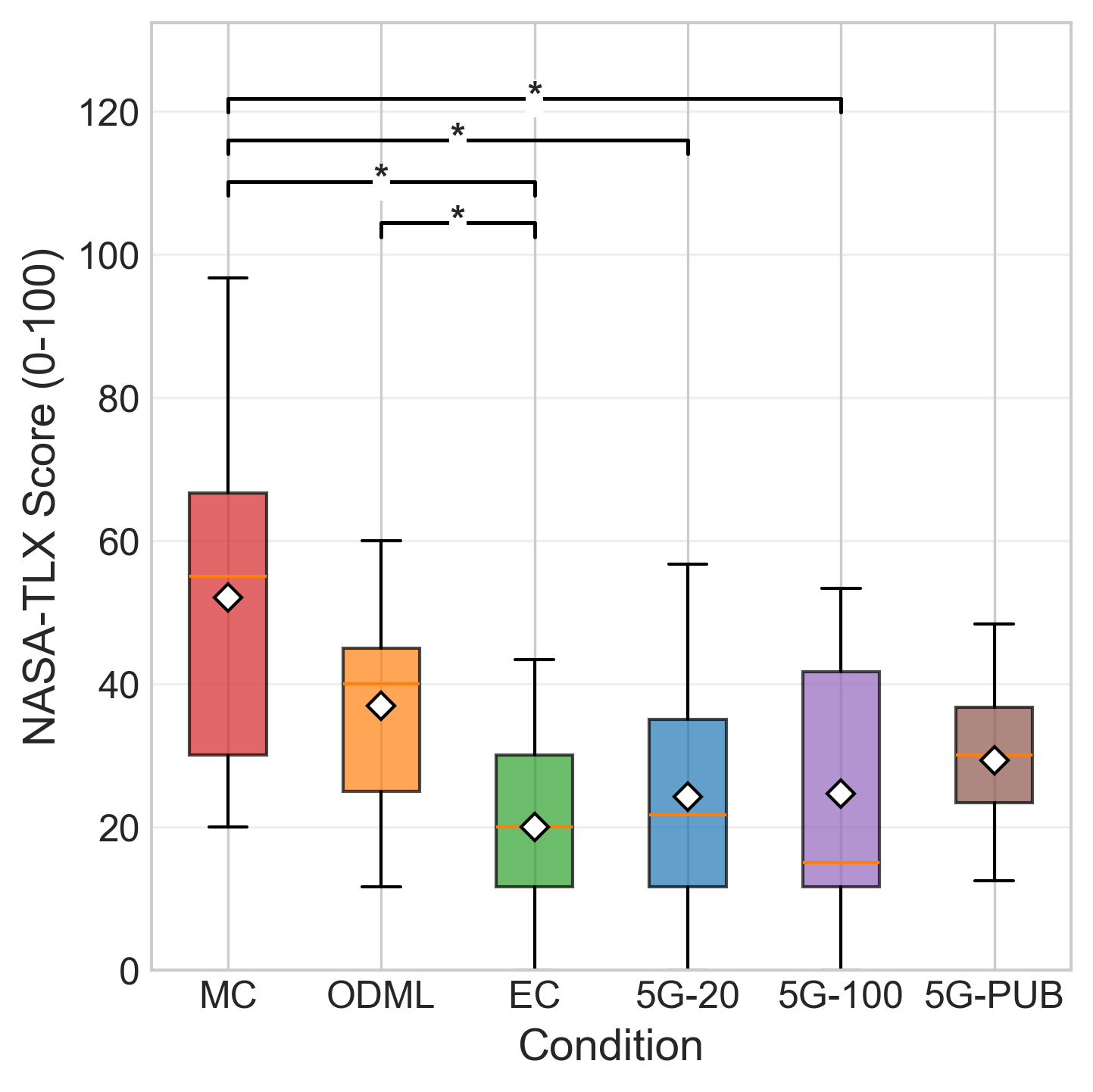}}}
\hfill
\subfloat[NASA-TLX Subscale Scores\label{fig:nasa_tlx}]{\includegraphics[width=0.45\linewidth]{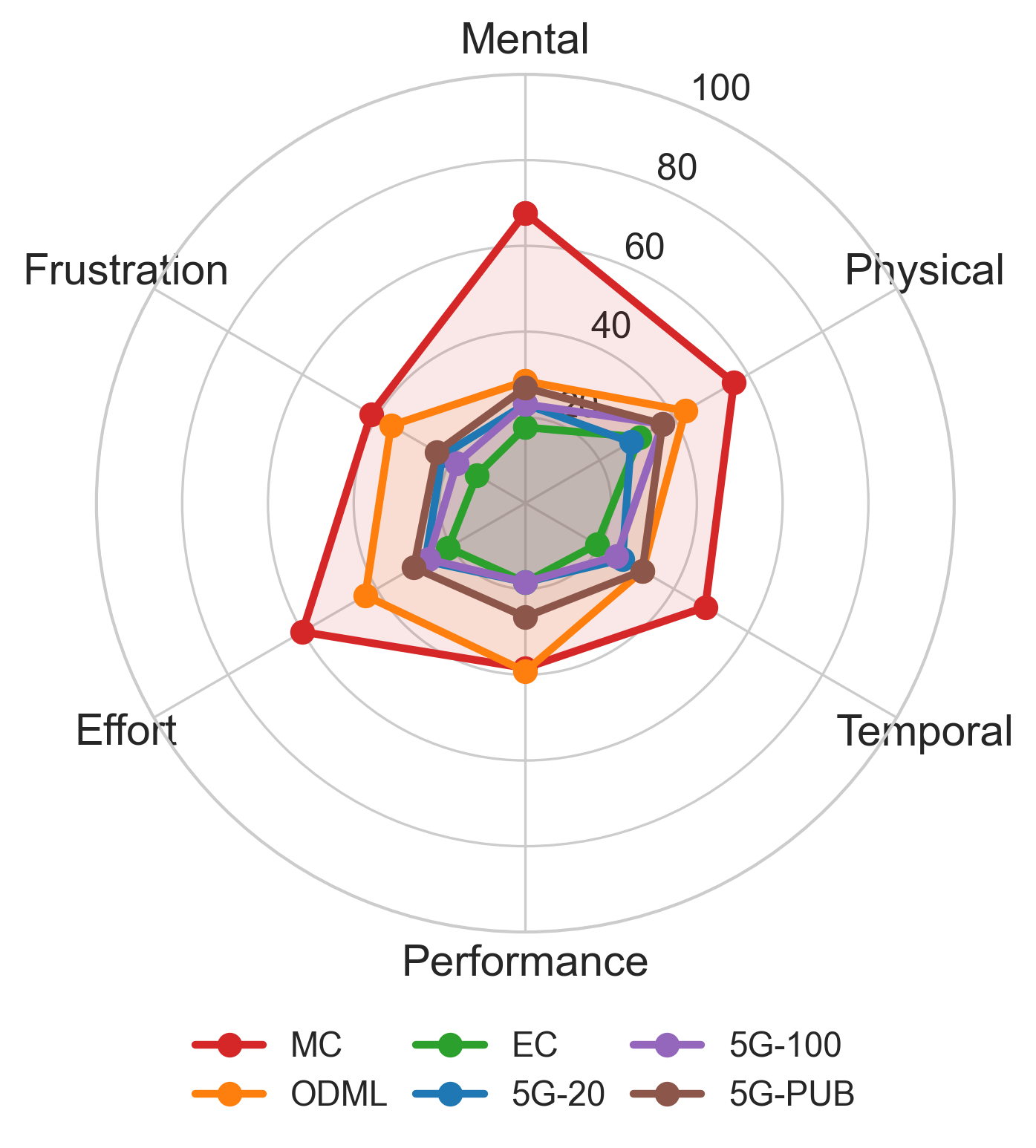}}
\caption{Subjective workload assessment: (a) NASA-TLX overall score per condition (box plot conventions as in Fig.~\ref{fig:results_objective}) and (b) NASA-TLX subscale scores shown as a polar plot. Manual control was characterized by the highest mental demand and effort, whereas the connected prosthesis conditions cluster near the center (low workload).}
\label{fig:results_subjective}
\end{figure*}

The assessment of subjective workload (Fig.~\ref{fig:results_tlx} and~\ref{fig:nasa_tlx}) reveals that manual control yielded the highest workload with a mean score of 52.1. This elevated workload was primarily driven by high mental demand (67.7) and effort (60.0), likely reflecting the cognitive burden associated with the explicit switching and sequential control of prosthesis functions.

On the other side, connected prosthesis conditions substantially reduced the workload relative to manual control. Ethernet achieved the lowest mean workload score (20.0), a 62\% reduction compared to manual control, while private 5G setups showed comparable scores (5G-20: 24.2, 5G-100: 24.6). The Friedman test confirmed a significant effect of condition on workload ($p < 0.0001$), and post-hoc Wilcoxon signed-rank tests showed that these reductions are statistically significant compared to manual control ($p_{adj} < 0.05$). Commercial 5G (29.3) showed a similar pattern but did not reach significance when compared to manual control ($p_{adj} = 0.056$).

On-device inference produced an intermediate workload (36.9), which was significantly higher than that obtained with Ethernet ($p_{adj} = 0.033$). The differences between ODML and the 5G conditions in overall workload did not reach statistical significance after correction. However, frustration was notably elevated for ODML (36.2), significantly higher than 5G-100 (18.5, $p_{adj} = 0.044$) and descriptively higher than EC (13.1, $p_{adj} = 0.067$). Participants reported missed detections and unresponsive controls.


\section{Discussion}

This study developed the first 5G-connected semi-autonomous prosthetic hand prototype and evaluated its performance under realistic deployment conditions. Our results demonstrate that offloading complex processing to MEC servers can support effective real-time semi-autonomous prosthesis control.  The performance of the developed smart connected prosthesis prototype matched that achieved with an ideal wired connection and significantly exceeded that obtained using both conventional manual control and on-device processing. Beyond objective task metrics, the findings further indicated that MEC-assisted shared control substantially reduced perceived cognitive workload, suggesting that the proposed framework can alleviate the user’s control burden in addition to accelerating computation. These observations are particularly relevant in light of the persistently high prosthesis abandonment rates, which are reported to exceed 40\% ~\cite{salminger_current_2022a} despite continuous mechatronic advancements. Demonstrating the feasibility and advantages of MEC-enabled prosthesis control paves the way towards a new generation of prosthetic systems enhanced with radically novel functions, including remote monitoring, data logging, large-scale learning and adaptation, and on-the-fly control updates, in addition to computational offloading ~\cite{chiariotti_future_2024}. The following discussion elaborates on the design, clinical, and deployment implications of the proposed MEC-enabled prosthetic systems.

\subsection{Feasibility of 5G-Connected Prosthetic Control}

The central finding is that 5G networks can support latency-critical prosthesis control in practice. All 5G configurations, including bandwidth-constrained 20~MHz and variable commercial networks, achieved similar performance to that obtained when using the wired Gigabit Ethernet in all outcome measures (task completion time, failure rate, and subjective workload). While the commercial 5G link exhibited the highest round-trip latency due to operation over a public network with uncontrolled bandwidth and multi-hop routing, it still significantly outperformed manual control in task completion time. There was no significant difference in performance between network conditions, and participants reported during debriefing that 5G and Ethernet configurations felt similarly effective.

The network latency ranged from 65~ms (private 5G 100~MHz link) to 234~ms (commercial 5G), which, compared to 10~ms for Ethernet, is a substantial increase in transmission delay. Yet, importantly, these differences did not translate into measurable performance degradation. The commonly cited 100--125~ms latency threshold for prosthetic control~\cite{farrell_optimal_2007} was established for direct proportional EMG, where user muscle signals continuously drive motor commands. Our shared-control architecture changes this latency-critical dynamic: since frames are processed continuously at 6--20~fps, the system pre-computes grasp solutions before the user initiates grasping. When the user triggers preshaping, the response uses an already-computed result rather than waiting for the full round trip of a new frame. Network latency is thus absorbed by the streaming pipeline rather than experienced as a response delay.

For clinical deployment, continuous streaming could be made more efficient by activating it only when needed. For example, an inertial measurement unit (IMU) on the prosthesis could detect reaching movements and trigger the start of the streaming, so that the vision pipeline is active during approach and grasp planning while remaining idle otherwise. This would preserve the benefits of pre-computed grasp solutions while substantially reducing network utilization and power consumption. Nevertheless, we used continuous streaming in this study to stress-test network capabilities and establish baseline 5G performance characteristics under sustained load.

\subsection{The Embedded Processing Bottleneck}
The on-device processing performed worst despite bypassing network latency, indicating the limitations of current embedded hardware for real-time prosthesis control using computationally demanding computer vision approaches. The Raspberry Pi 5 achieved only 3~fps, compared with 6--20~fps for the network-offloaded conditions (Table~\ref{tab:network}). The 429~ms processing time meant that grasp solutions were frequently stale or unavailable when users triggered preshaping.

The 75.9\% failure rate for ODML did not reflect model accuracy; both configurations used the same YOLOv11 architecture, 
with the on-device condition using the nano variant and the server using the large variant. The problem was temporal: at 3~fps, valid detections were often unavailable when users initiated grasps. Users reported repeatedly triggering without response, leading to frustration and disengagement.

This finding has implications for prosthetic system design. Despite the appeal of fully self-contained wearable systems, current embedded hardware cannot support real-time deep learning perception at useful frame rates. The order-of-magnitude performance gap between embedded CPUs and discrete GPUs for inference suggests that network offloading will remain advantageous, at least until embedded AI accelerators are improved substantially.

\subsection{Comparison with Prior Work}

Our task completion times (8.6~s for connected prosthesis conditions) align with Mouchoux et al.'s MYO-PACE system~\cite{mouchoux_artificial_2021}, which achieved 9--11~s using embedded processing with continuous sensor fusion. This similarity suggests that our network-offloaded architecture achieves comparable performance while enabling more computationally intensive vision models than embedded processors can support. Indeed, in ~\cite{mouchoux_artificial_2021}, the processing was fully implemented on a dedicated lab computer using a tethered configuration. 

Shatilov et al.~\cite{shatilov_using_2019} achieved 90--370~ms delays for EMG gesture classification offloaded through a smartphone intermediary. Our architecture differs in three important ways: (1) direct 5G connection using the full 5G setup with MEC enabled base station with inference server, without smartphone hop, reducing potential latency and failure points; (2) vision-based grasp planning to implement semi-automatic control rather than discrete gesture classification (conventional manual control); and (3) evaluation on clinical-grade prosthesis hardware with real grasping tasks rather than assessing the classification accuracy alone. Our total latencies (35--270~ms depending on condition) fall within their reported range despite substantially more complex processing (object detection, segmentation, PCA measurement, grasp planning vs.\ EMG classification).

The vision pipeline we implemented is straightforward: YOLO detection, PCA-based measurement, and rule-based grasp selection. This simplicity was deliberate, since the main contribution of the present work is the development and assessment of a connected system architecture, rather than the development of a particular approach to grasp planning, machine learning, and computer vision. Importantly, the proposed framework supports more sophisticated approaches, such as learned grasp planners~\cite{newbury_deep_2023}, dexterous multi-finger coordination~\cite{andrychowicz_learning_2020}, and multimodal sensor fusion combining vision with gaze and EMG~\cite{cognolato_multimodal_2022}, without requiring structural changes. This also includes methods that are not necessarily related to semi-autonomous control, but are still too complex to be run locally [], such as the use of large-scale high-fidelity musculoskeletal modeling [] and high-density EMG decomposition to decode user motion intention for prosthesis control []. The only modification necessary to integrate any of these methods would be to implement them on the inference server and transmit the sensor data (e.g., high-density EMG in addition to or instead of the image stream). This is a unique and powerful feature of the connected prosthesis concept we are proposing: the control approach can be changed at the Edge and/or Cloud without any modifications to the rest of the system or user interface. This modularity is a key advantage of edge offloading; as vision algorithms improve, prosthesis users benefit without hardware upgrades.

\subsection{Deployment Considerations and Practical Constraints}

Our results support the deployment of 5G-connected prosthetic systems outside controlled laboratory settings. Several practical factors merit discussion.

\emph{Portable operation:} The control unit (Raspberry Pi~5, 5G modem, and camera) is powered by a 20,000~mAh (74~Wh) USB power bank, providing approximately 5--6~hours of continuous operation. During development, the system was successfully operated while the user moved in the lab, confirming basic portability. Further miniaturization is expected as 5G modem form factors shrink and single-board computers become more compact.

\emph{Network performance:} The commercial 5G exhibited higher latency variability (P95 = 890~ms), in comparison to private networks (P95 = 287--472~ms) reflecting real-world congestion and routing. Despite this variability, the task performance remained functional. Coverage availability likely matters more than peak performance for practical deployment. Beyond raw latency and throughput, 5G networks provide standardized mechanisms that can be leveraged to improve service robustness in real-world conditions. In particular, network slicing~\cite{foukas_network_2017} can dedicate a virtual network partition with guaranteed quality of service to prosthetic data streams, while multi-connectivity~\cite{agiwal_survey_2021} can maintain simultaneous links to multiple cells to improve reliability during handovers or congestion. Such mechanisms are especially relevant for uplink-heavy perception streams and latency-sensitive downlink control messages.

\emph{Graceful degradation:} Our prototype does not currently implement fallback to manual control during network outages. Production systems should detect connectivity loss and seamlessly transition to EMG-only operation, since network conditions vary by location and time.

\emph{Privacy:} Streaming RGB-D imagery raises privacy concerns, as the camera captures not only target objects but also surrounding scenes. Healthcare deployments would require encrypted connections, potentially on-device anonymization (blurring non-target regions), or edge computing within institutional networks where data remains local.

\emph{Cost model:} Cellular data plans add ongoing cost beyond prosthesis hardware. Private 5G networks offer lower latency but require institutional investment. Commercial 5G is more accessible but subject to carrier pricing and coverage gaps. The cost-benefit tradeoff warrants further analysis, including potential healthcare-specific data arrangements.

\subsection{Limitations}

\emph{Network environment:} Testing occurred in a controlled laboratory with line-of-sight to the 5G base station. Real-world 5G performance varies with congestion, signal strength, building penetration, and handovers. Although the commercial 5G condition partially captures real-world variability, further evaluation under diverse environmental and mobility conditions is required.

\emph{Participant population:} Able-bodied participants using a prosthesis simulator do not fully represent prosthesis users. However, the aim of the present study was to demonstrate the concept of MEC-enabled prosthesis control and test its technical feasibility, whereas the future work will explore the clinical translation. The consistent performance improvements across all 13 participants suggest the benefits of 5G offloading would possibly generalize to clinical users. One advantage of semi-automatic control is that it is less dependent on the user's skill, as the most complex system functions are controlled automatically, while the manual commands that the user needs to generate are simple (e.g., triggering automatic preshape and closing the hand).  

\section{Conclusion And Future Work}

This paper presented the first prototype of a 5G-connected semi-autonomous prosthetic hand and provided a systematic experimental evaluation of its performance under different connectivity configurations. The results demonstrated that the performance of online prosthesis control using offloading of computationally demanding computer vision processing over 5G networks matches that achieved when using wired Ethernet connectivity. Private 5G (20 and 100~MHz), commercial 5G, and Ethernet configurations achieved equivalent task completion time, failure rate, and subjective workload. Further, the connected prosthesis significantly outperformed conventional manual control (clinical standard). These findings establish 5G edge/cloud offloading as a practical path to deploying sophisticated compute-intensive control methods in prosthetic systems, effectively overcoming the barrier of embedded processing constraints. Looking forward, the following directions can be considered for future work:

\emph{Hybrid architectures:} Running lightweight detection on-device while offloading grasp planning to the edge could reduce latency while preserving model capability and enabling graceful degradation in case of loss of connection. This split requires careful pipeline partitioning.

\emph{Predictive control:} Tracking arm movement and gaze direction could anticipate grasp targets and pre-compute solutions, further masking network latency.

\emph{Clinical validation:} Longitudinal studies with prosthesis users in everyday environments are needed to validate real-world applicability, measuring not only task performance but also adoption, satisfaction, and quality-of-life outcomes.
\section*{Acknowledgments}
The authors thank Alexander Løvig Borg for 3D-printed mounting components and logistical support, and Martin Alexander Garenfeld for initial myoelectric control implementation. We thank all participants for their time and effort.


\bibliographystyle{IEEEtran}
\bibliography{references} 

\end{document}